\documentclass{elsart}

\usepackage{natbib}
\usepackage{amssymb}

\usepackage{graphicx}
\usepackage{dcolumn}
\usepackage{bm}

\def\bea{\begin{eqnarray}}
\def\eea{\end{eqnarray}}

\begin{document}

\begin{frontmatter}


\title{Autocorrelations from the scale dependence of transverse-momentum fluctuations in Hijing-simulated Au-Au collisions at $\sqrt{s_{NN}} = $ 200 GeV}

\author{Qingjun Liu,\thanksref{liu}} \author{Duncan J. Prindle and Thomas A. Trainor}

\thanks[liu]{On leave from Chinese Academy of Sciences, P.R. China}

\address{CENPA, Box 354290, University of Washington, Seattle, WA 98195}


\begin{abstract}
We present measurements of the scale (bin-size) dependence of event-wise mean transverse momentum $\langle p_{t} \rangle$ fluctuations for Au-Au collisions at $\sqrt{s_{NN}} = 200$ GeV simulated with the Hijing-1.37 Monte Carlo. We invert those scale distributions to obtain autocorrelation distributions on pseudorapidity $\eta$ and azimuth $\phi$ difference variables $(\eta_\Delta,\phi_\Delta)$. The autocorrelations have a simple structure dominated by a same-side (on azimuth) peak with similar widths on the two difference variables. With jet quenching imposed, the same-side peak widths increase slightly in both directions, and the amplitude is substantially reduced. The same-side peaks are attributed to minijets, observed in this study as local {\em velocity} correlations of charged particles with $p_t < $ 2 GeV/c.  
\end{abstract}

\begin{keyword}
Mean-$p_{t}$ fluctuations \sep $p_t$ correlations \sep heavy ion collisions \sep scale dependence \sep inverse problem

\PACS 24.60.Ky \sep 25.75.Gz

\end{keyword}

\end{frontmatter}


\section{Introduction}

According to QCD theory a phase transition from hadronic nuclear matter to a color-deconfined medium (quark-gluon plasma or QGP) may occur in relativistic heavy ion collisions at RHIC~\cite{QCD}. Measurements of nonstatistical fluctuations of multiplicity $n$ and event-wise mean transverse momentum $\langle p_t \rangle$ have been proposed to identify and study the QGP~\cite{Rstock,Gbaym,Stodolsky,Shuryak,SRS,Berdnikov,droplets}. Nonstatistical fluctuations could also result from minijet production \cite{QT} or from other dynamical processes. Minijets (hadron fragments from a {\em minimum bias} parton distribution) should be copiously produced in the initial stage of ultra-relativistic heavy ion (HI) collisions at RHIC \cite{pundits}. A colored medium formed in the prehadronic phase of such collisions could cause jets/minijets to lose energy through induced gluon radiation \cite{Quench}, a dissipative process conventionally called jet quenching in the case of higher-$p_t$ partons. The properties of the dissipative medium would then determine specific energy loss \cite{QuenDen} and other aspects of minijet modification which could be studied with fluctuation measurements. 

In a previous paper~\cite{QT} we reported an analysis of $\langle p_t \rangle$ fluctuations in simulated Au-Au collisions at 200 GeV at a fixed {\em scale} (bin size), using Hijing-1.37 \cite{Hijing} to study jet/minijet contributions to those fluctuations. According to that study minijets are the dominant source of nonstatistical $\langle p_t \rangle$ fluctuations in heavy ion collisions as modeled by Hijing. The $\langle p_t \rangle$ fluctuation amplitude quantitatively measures the extent of jet production and jet quenching in that model. A measurement of nonstatistical $\langle p_t \rangle$ fluctuations in Au-Au collisions at 130 GeV by the STAR collaboration~\cite{ptprl} was also made at a single scale (the ($\eta,\phi$) detector acceptance). Comparison of that STAR measurement to our Hijing analysis has been inconclusive concerning minijets as the fluctuation source. 

In this Letter we extend our fixed-scale Hijing analysis to determine the {\em scale dependence} of $\langle p_t \rangle$ fluctuations. By inverting that scale dependence with a standard numerical technique~\cite{inverse} we obtain two-particle correlations induced by minijets in the form of {\em autocorrelation} distributions. This more differential study of fluctuations and correlations provides detailed information about parton dissipation and resulting minijet structure in the Hijing model and establishes a reference for similar studies of RHIC data.

\section{Relating Fluctuations and Autocorrelations}

The {\em scale dependence} of event-wise $\langle p_t \rangle$ fluctuations on kinematic variables ($\eta, \phi$) is determined by the structure of underlying two-particle correlations~\cite{cltps}. Variation of nonstatistical fluctuations (measured by variance differences) with bin size or scale can be inverted to obtain an {\em autocorrelation} distribution on difference variables. 
The autocorrelation distribution, a projection by averaging of a two-particle distribution on a momentum space [{\em e.g.,} $(\eta_1,\eta_2)$] onto its difference variable(s) ({\em e.g.,} $\eta_\Delta \equiv \eta_1 - \eta_2$), is often simply and directly interpretable in terms of physical phenomena. The actual fluctuation analysis is carried out on two-dimensional (2D) space $(\eta,\phi)$ within detector acceptance $(\Delta \eta,\Delta \phi)$ divided into $M$ macrobins of variable size (scale) $(\delta \eta,\delta \phi)$ which are in turn divided into $m_\delta$ microbins of fixed size $(\epsilon_\eta,\epsilon_\phi)$ (microbins are used in the numerical integration). We restrict to 1D space $x$ ($x = \eta,\, \phi$) to summarize the method and then generalize to 2D $(\eta,\phi)$ for the analysis. 

A 1D macrobin of size $\delta x$ contains integrated $p_t(\delta x)$ (a scalar sum over particles in the bin) and particle multiplicity $n(\delta x)$. To measure fluctuations of $p_t$ {\em relative to} particle multiplicity $n$ we employ $\{p_t(\delta x) - n(\delta x)\, \hat p_t\} / \sqrt{\bar n(\delta x)}$ as the basic statistical quantity and define the scale-dependent {\em per-particle conditional variance} of $p_t$ {\em given} multiplicity $n$ by $\sigma_{p_t:n}^2(\delta x) \equiv \overline{\{p_t(\delta x)-n(\delta x)\, \hat p_t\}^2}/\bar n(\delta x)$, where $\hat p_t\equiv \bar p_t(\Delta x) / \bar n(\Delta x)$ is the inclusive mean per-particle $p_t$ and the overline represents an average over all bins of size $\delta x$ in all events~\cite{QT,ptprl}. 

We relate that variance to an equivalent {\em autocorrelation} distribution on difference variable $x_\Delta = x_1 - x_2$ in two-particle space $(x_1,x_2)$ by the following argument. Each autocorrelation element $A_{k}(\epsilon_x)$ is an average over the $k^{th}$ diagonal of 2D microbins in a 2D macrobin on $(x_1,x_2)$ and a further average over all macrobins in all events~\cite{inverse}. Given autocorrelations for object and reference distributions we define the {\em difference} autocorrelation  $\Delta A_k \equiv A_{k,obj} - A_{k,ref}$. Since the content of a 1D macrobin on $x$ is also equal to a sum over the $m_\delta$ 1D microbins within it, we can relate the variance to a difference autocorrelation with an integral equation expressed as a discrete sum over microbins 
\bea \label{eq1}
&& \sigma^2_{p_t:n}(m_\delta \epsilon_x) = \overline{\{p_t(m_\delta \epsilon_x)-n(m_\delta \epsilon_x)\, \hat p_t\}^2}/\bar n(m_\delta \epsilon_x)\\ \nonumber
 &&= \sum_{a,b=1}^{m_\delta} \overline{ \frac{\left\{p_t(\epsilon_x)-n(\epsilon_x)\, \hat p_t\right\}_a \cdot\left\{p_t(\epsilon_x)-n(\epsilon_x)\, \hat p_t\right\}_b}{{m_\delta \,\bar n(\epsilon_x)}}} \\ \nonumber
&&= \sum_{k=1-m_\delta}^{m_\delta -1} \hspace{-.1in} K_{m_\delta;k} \frac{\bar n_k}{\bar n} \sum_{1 \leq a,b \leq m_\delta}^{a-b = k}\frac{1}{m_\delta - |k|} \frac{\overline{ \left\{\cdots \right\}_a \cdot \left\{\cdots \right\}_b}}{\sqrt{\bar n_a} \, \sqrt{\bar n_b}} \\ \nonumber
 && \equiv  \sum_{k=1-m_\delta}^{m_\delta -1}  K_{m_\delta;k}  \frac{\bar n_k}{\bar n}\frac{\Delta A_k(p_t:n;\epsilon_x)}{\sqrt{A_{k,ref}(n;\epsilon_x)}},
\eea
where the integration kernel $K_{m_\delta;k} \equiv (m_\delta - |k|)/m_\delta$, and the overline indicates an average over all macrobins of size $\delta x$ in all events. To simplify summations we shift the binning on $x_\Delta$ by 1/2 bin and assume symmetry about the origin on $x_\Delta = (k - 1/2) \epsilon_x$, with $k \in [1,m_\delta]$. The kernel then becomes $K_{m_\delta;k} = (m_\delta - k+1/2)/m_\delta$. A reference {\em number} autocorrelation is defined by $A_{k,ref}(n,\epsilon_x)  \equiv \bar n_k^2(\epsilon_x)\equiv \overline{\bar n_a(\epsilon_x) \bar n_b(\epsilon_x)}_{a-b=k}$ (a microbin average on the $k^{th}$ diagonal). 

The per-hadron conditional {\em variance difference}
$\Delta \sigma^2_{p_t:n}(\delta x) \equiv \sigma_{p_t:n}^2(\delta x) - \sigma_{\hat p_t}^2$ integrates two-particle correlations over the scale interval $[0,\delta x]$~\cite{cltps}. The inclusive single-particle variance $\sigma_{\hat p_t}^2$ corresponds to self pairs in two-particle sums. If the self pairs are excluded from autocorrelation definitions then $ \sigma^2_{p_t:n} \rightarrow \Delta \sigma^2_{p_t:n}$ in the above equations. To obtain a correlation measure which is independent of the microbin size we define autocorrelation densities by $\epsilon_x^2\, \Delta \rho(p_t:n,k\, \epsilon_x) \equiv \bar n_k / \bar n \cdot \Delta A_k(p_t:n;\epsilon_x)$ and $\epsilon_x^2\, \rho_{ref}(n,k\, \epsilon_x) \equiv A_{k,ref}(n;\epsilon_x)$ (we absorb the $O(1)$ multiplicity-ratio factor into the definition of the difference density). We then have
\bea \label{eq2}
\Delta \sigma^2_{p_t:n}(m_\delta \, \epsilon_x) &=& 2\, \sum_{k=1}^{m_\delta} \epsilon_x\,  K_{m;k} \,  \frac{\Delta \rho(p_t:n,k\, \epsilon_x)}{\sqrt{\rho_{ref}(n,k\, \epsilon_x)}},
\eea
where $\Delta \rho / \sqrt{\rho_{ref}}$ is the desired intensive, per-particle measure of two-particle correlations projected onto difference variable $x_\Delta$.

For the 2D scale-dependence analysis we generalize $\delta x \rightarrow  (\delta \eta,\delta \phi)$ to obtain the per-particle conditional $p_t$ variance difference $\Delta \sigma^2_{p_t:n}(\delta \eta,\delta \phi)$ in the discrete form 
\bea \label{inverse}
\Delta \sigma^2_{p_t:n}(m_\delta \, \epsilon_\eta, n_\delta \, \epsilon_\phi) &\equiv& 2 \sigma_{\hat p_t} \Delta \sigma_{p_t:n}(m_\delta \, \epsilon_\eta, n_\delta \, \epsilon_\phi) \\ \nonumber
= 4 \sum_{k,l=1}^{m_\delta,n_\delta} \epsilon_\eta \epsilon_\phi & &\hspace{-.25in} K_{m_\delta n_\delta;kl}  \,  \frac{\Delta \rho(p_t:n;k\, \epsilon_\eta,l\, \epsilon_\phi)}{\sqrt{\rho_{ref}(n;k\, \epsilon_\eta,l\, \epsilon_\phi)}},
\end{eqnarray}
with 2D kernel $K_{m_\delta n_\delta;kl} \equiv (m_\delta - {k + 1/2})/{m_\delta} \cdot (n_\delta-{l+1/2})/{n_\delta}$ representing the 2D binning process. This Fredholm equation can be inverted to obtain autocorrelations in the form $\Delta \rho / \sqrt{\rho_{ref}}$ from $\langle p_t \rangle$ fluctuation scale dependence on $(\delta \eta,\delta \phi)$ \cite{inverse}.

\section{Hijing Simulations}

We used event generator Hijing-1.37 to produce minimum-bias event ensembles with $\sim $ 1M total events of two types of Monte Carlo collision: 1) {\em quench-off}\, Hijing -- jet production enabled but not jet quenching  and 2) {\em default} or {\em quench-on} Hijing -- jet production and jet quenching enabled. Study of those event types should help to clarify the role of jet production and jet quenching in the correlation structure of RHIC heavy-ion collisions. For this analysis charged particles with pseudorapidity $|\eta| < $ 1, transverse momentum $p_t \in [0.15,2]$ GeV/c and full azimuth were accepted. Minimum-bias Hijing events were classified according to impact parameter $b$ (fm) to estimate collision centrality. Six centrality classes were defined: 80-95\%, 65-80\%, 50-65\%, 30-50\%, 15-35\% and 0-15\% of the total cross section.

\section{Fluctuation Scale Dependence}

\begin{figure}[h]
\begin{tabular}{cc}
\includegraphics[keepaspectratio,width=1.65in]{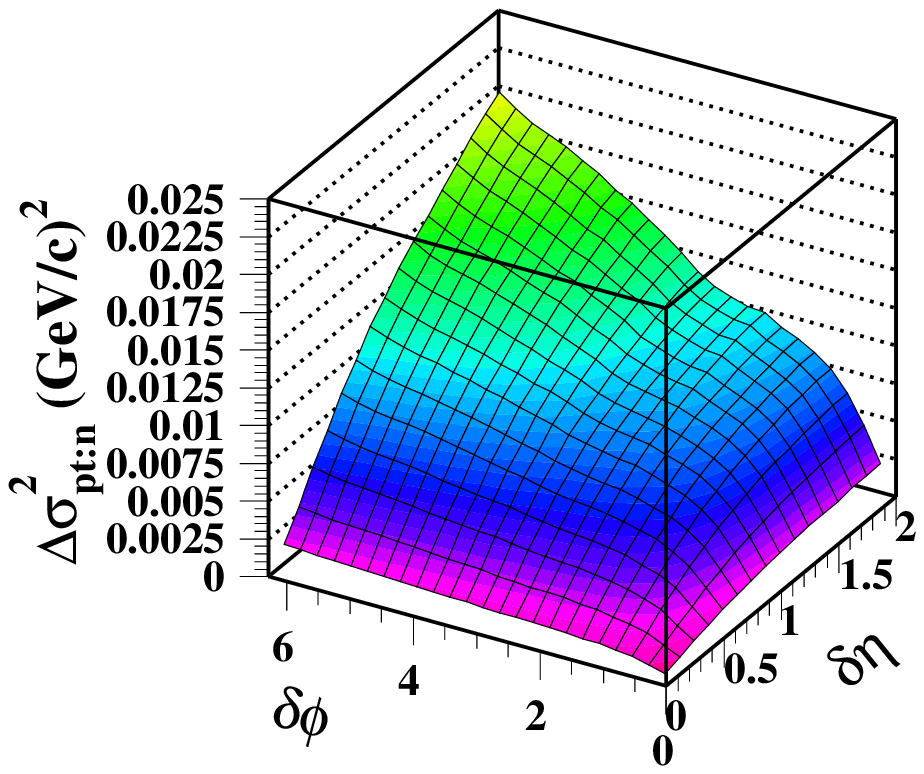} &
\includegraphics[keepaspectratio,width=1.65in]{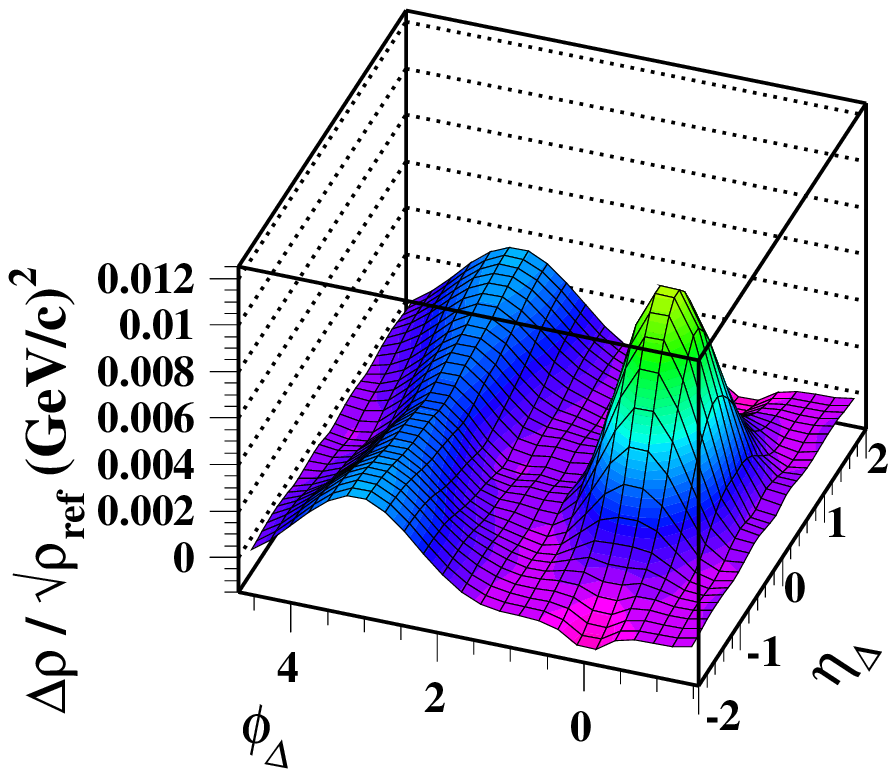} \\
\includegraphics[keepaspectratio,width=1.65in]{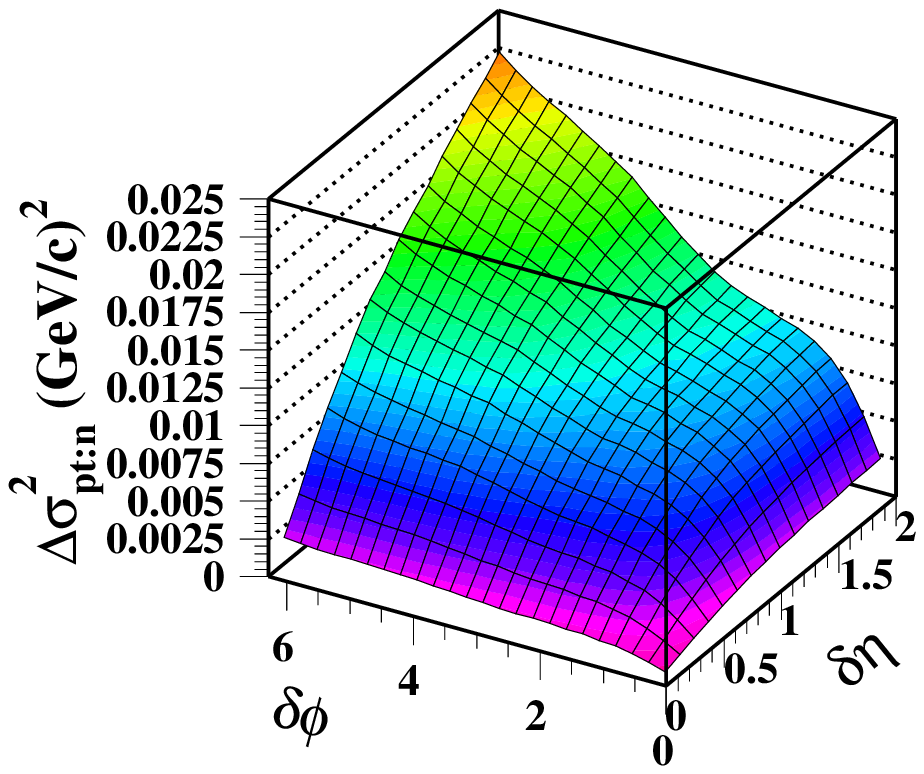} &
\includegraphics[keepaspectratio,width=1.65in]{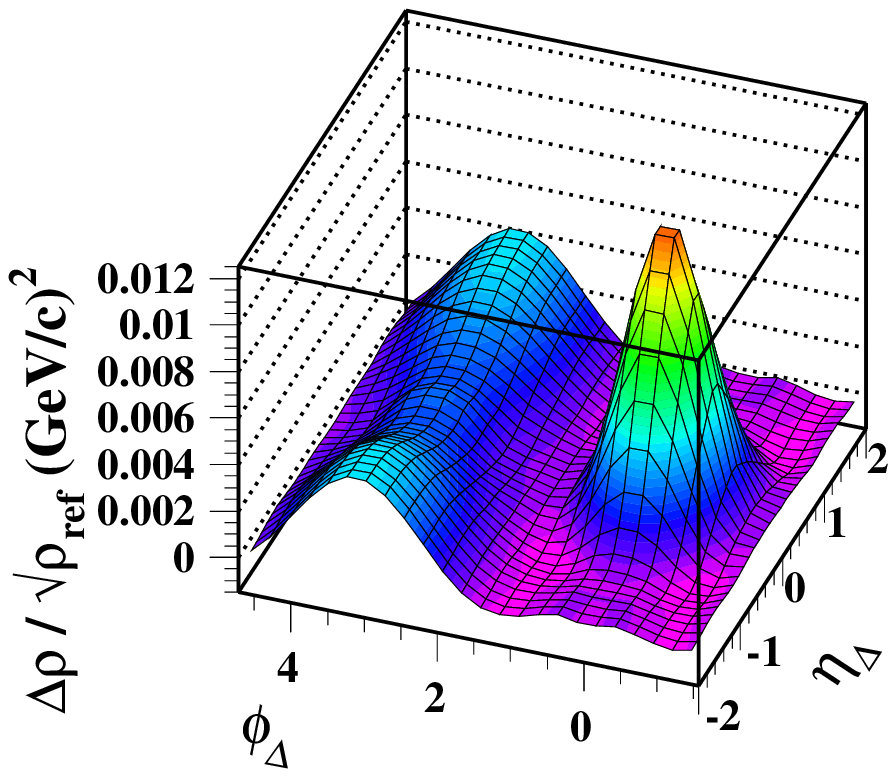}
\end{tabular}
\caption{Left panels: Variance difference $\Delta \sigma^2_{p_{t}:n}$ (GeV/c)$^2$ on scales $(\delta \eta,\delta \phi)$ for 65-80\% central (upper panel) and 15-30\% central (lower panel) Hijing events with jet quenching off. Right panels: Corresponding autocorrelations $\Delta \rho / \sqrt{\rho_{ref}}$ (GeV/c)$^2$ on difference variables $(\eta_\Delta,\phi_\Delta)$.\label{fig1}}
\end{figure}

In Fig.~\ref{fig1} (left panels) we present variance differences $\Delta \sigma^2_{p_t:n}(\delta \eta,\delta \phi)$ (GeV/c)$^2$ for Hijing quench-off collisions in centrality classes 65-80\% and 15-30\%. The variance differences typically increase monotonically with pseudorapidity scale $(\delta\eta)$ but are more complex on azimuth scale $(\delta\phi$). While fluctuation measurements can certainly be compared to theoretical predictions, the underlying physical mechanisms are often more evident in the corresponding autocorrelation distributions.

\section{Extraction of Autocorrelations}

\begin{figure}[h]
\begin{tabular}{cc}
\begin{minipage}{.47\linewidth}
\includegraphics[keepaspectratio,width=1.65in]{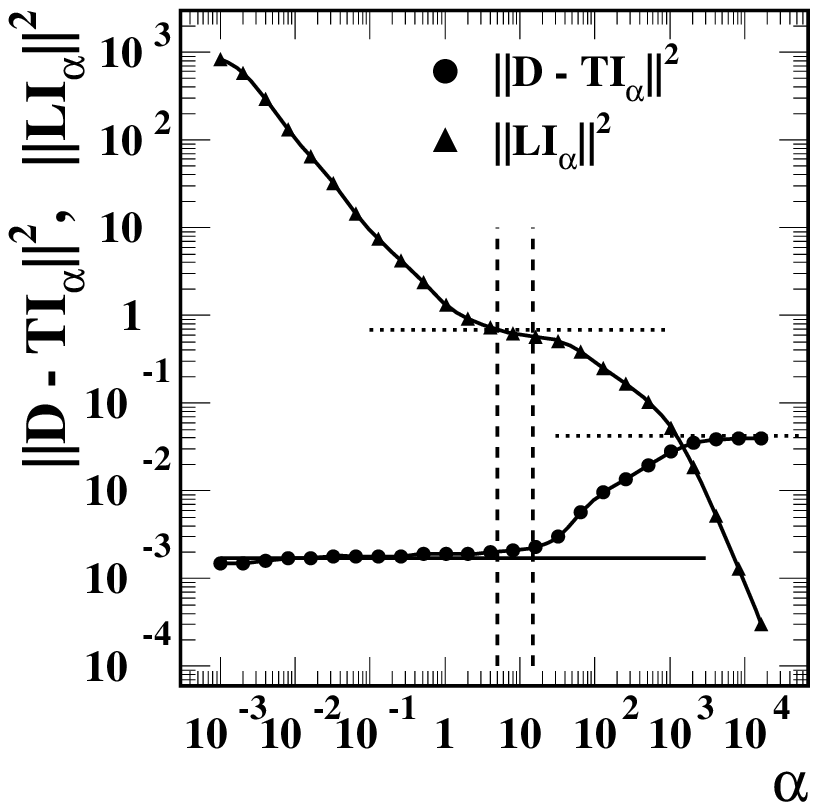}
\end{minipage} &
\begin{minipage}{.47\linewidth}
\includegraphics[keepaspectratio,width=1.65in]{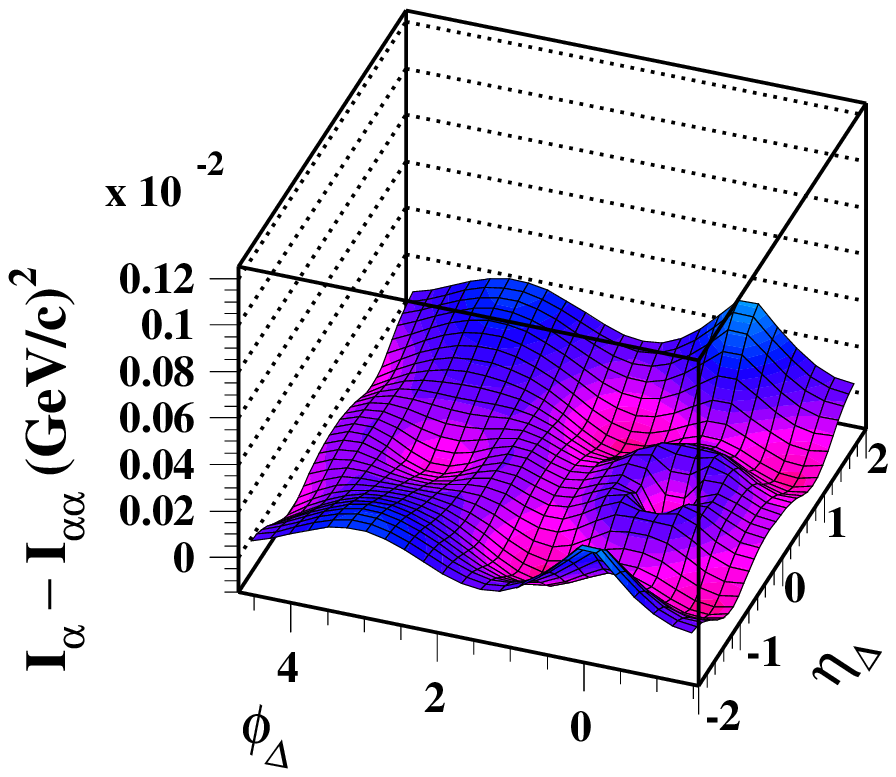}
\end{minipage}\\
\end{tabular}
\caption{Left panel: Distribution of $||{\bf D - T I}_\alpha||^2$ (dots) and $||{\bf L I}_\alpha||^2$ (triangles) $vs$ $\alpha$ for the 80-95\% central quench-on data in Fig.~\ref{fig2} (upper-right panel). Deviation of $||{\bf D - T I}_\alpha||^2$ from the extrapolation (horizontal solid line) indicates systematic error due to image smoothing. The vertical dashed lines bracket the optimal choice of $\alpha$. Right panel: Distribution of estimated smoothing error ${\bf I}_{\alpha} - {\bf I}_{\alpha\alpha}$ on $(\eta_\Delta,\phi_\Delta)$, also for the 80-95\% central autocorrelation distribution in Fig.~\ref{fig2}. Note the 10$\times $ scale reduction compared to Fig.~\ref{fig2}. \label{fig3}}
\end{figure}

Fig.~\ref{fig1} (right panels) shows 2D autocorrelation distributions inferred from the fluctuation scale dependence in the left panels by inverting Eq.~(\ref{inverse}) \cite{inverse}. We designate the variance difference distributions on the left as data ${\bf D} = \Delta \sigma^2_{p_t:n}$ and the autocorrelations on the right as image ${\bf I} = {\Delta \rho(p_t:n)}/{\sqrt{\rho_{ref}(n)}}$, related by Eq.~(\ref{inverse}) in the form of matrix equation ${\bf D = T\, I+N}$, where {\bf N} is the statistical `noise' on the measured data. The image can be inferred in principle by the simple matrix manipulation ${\bf T}^{-1}{\bf D = I +}{\bf T}^{-1}{\bf N}$. That approach results in an image dominated by small-wavelength noise in ${\bf T}^{-1}{\bf N}$, since the inverse matrix ${\bf T}^{-1}$, essentially a differentiation, acts as a `high-pass' filter to amplify the small-wavelength components of {\bf N}. 
To control statistical noise on the image a standard {\em regularization} procedure is employed~\cite{inverse}. We treat the bins of image hypothesis ${\bf I}_\alpha$ as free parameters in a $\chi^2$ fit and add a {\em smoothing} term with Lagrange multiplier $\alpha$ to obtain $\chi^2_\alpha \equiv ||{\bf D - T I}_\alpha||^2 + \alpha ||{\bf L I}_\alpha||^2$. The first term measures the deviation of the integrated image from the data, and the second term, with linear operator ${\bf L}$, measures small-wavelength structure on the image. The optimum choice of $\alpha$ maximally reduces small-wavelength noise while minimally distorting meaningful structure on the image. Distributions of $||{\bf D - T I}_\alpha||^2$ and $||{\bf L I}_\alpha||^2$ on $\log \alpha$, as shown in Fig.~\ref{fig3} (left panel), are used to determine the optimum $\alpha$ value, bracketed in that case by the vertical dashed lines.

\section{Results}

\begin{figure}[h]
\begin{tabular}{cc}
\includegraphics[keepaspectratio,width=1.65in]{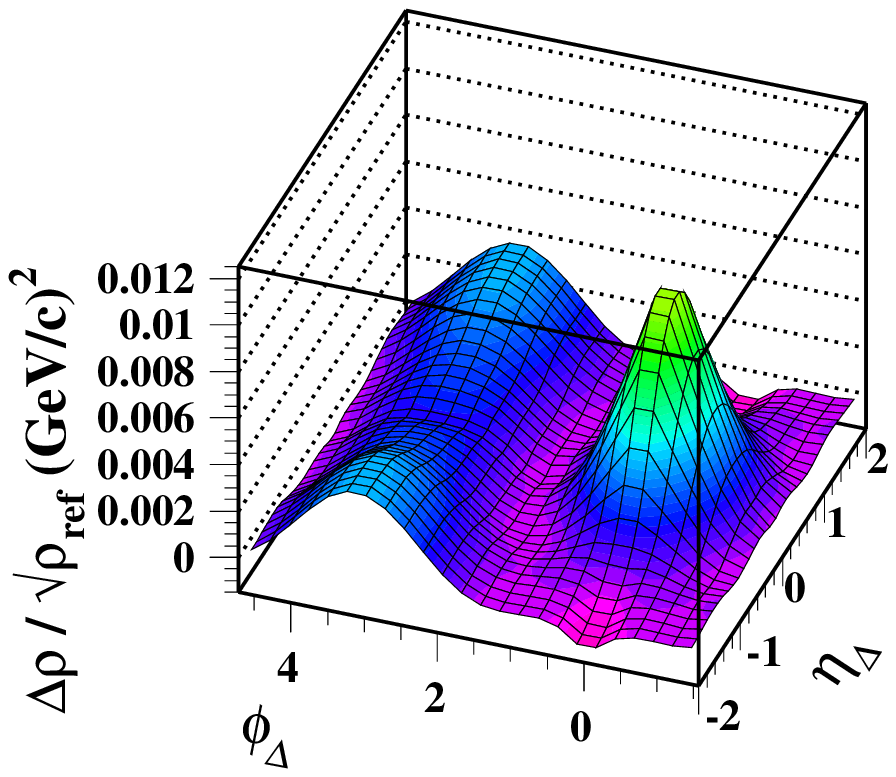} &
\includegraphics[keepaspectratio,width=1.65in]{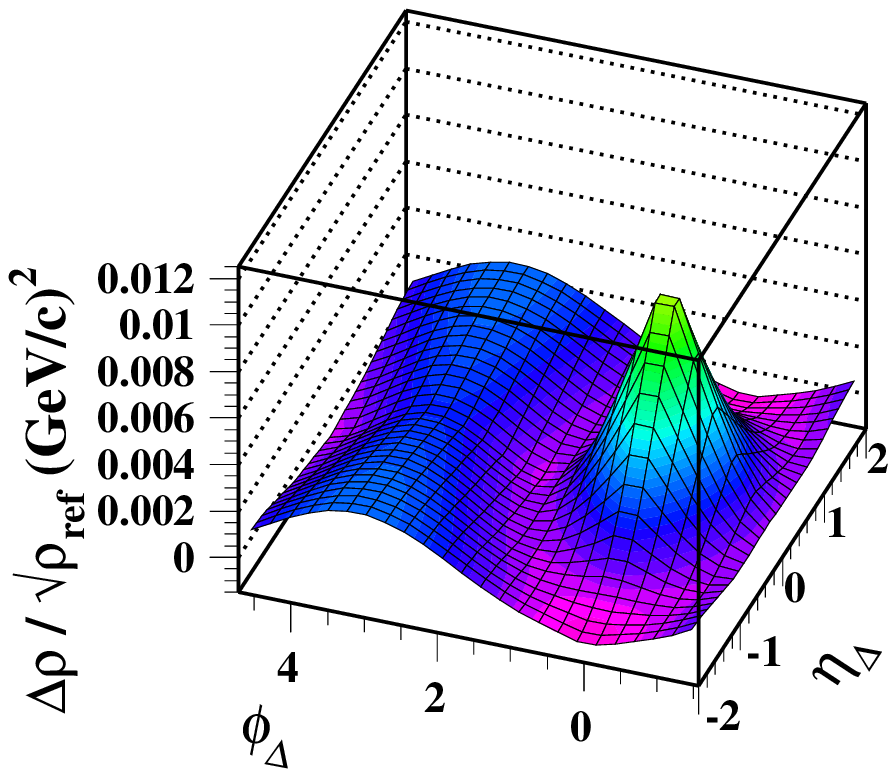} \\
\includegraphics[keepaspectratio,width=1.65in]{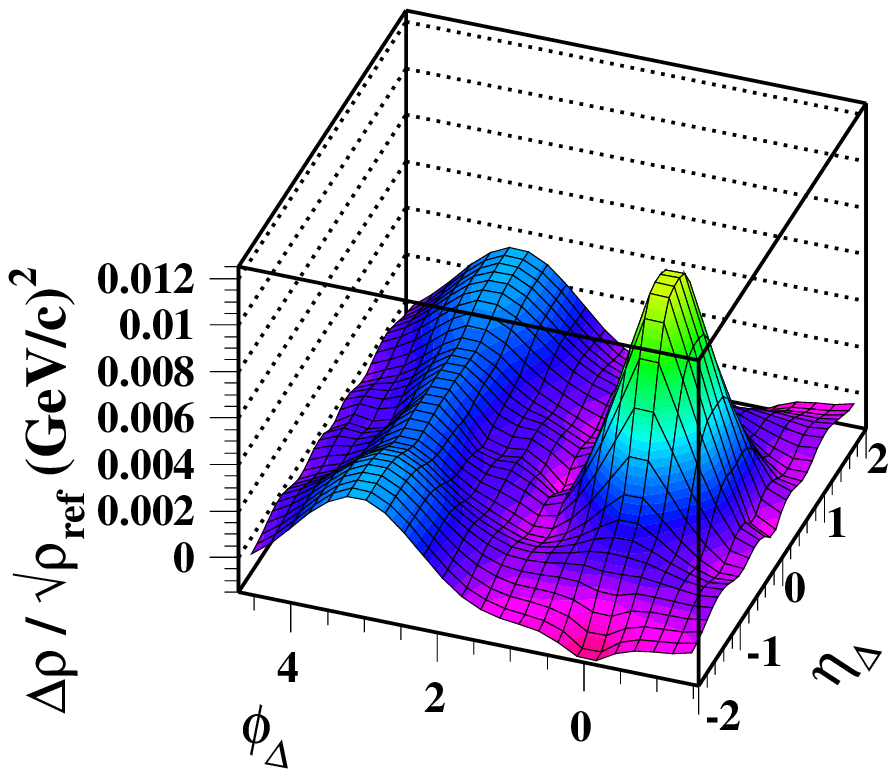} &
\includegraphics[keepaspectratio,width=1.65in]{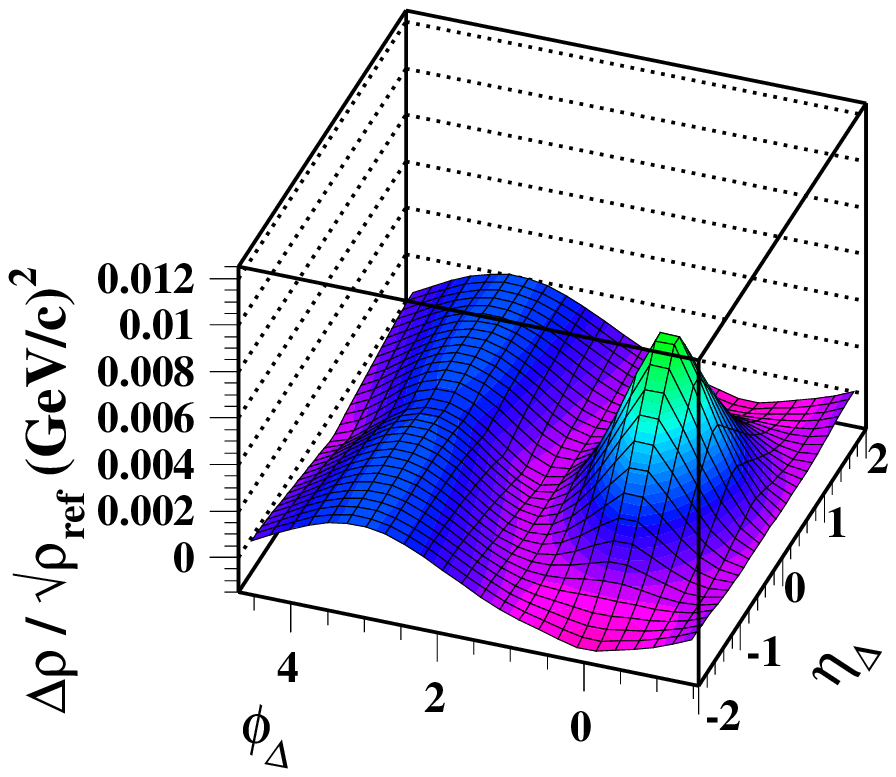} \\
\includegraphics[keepaspectratio,width=1.65in]{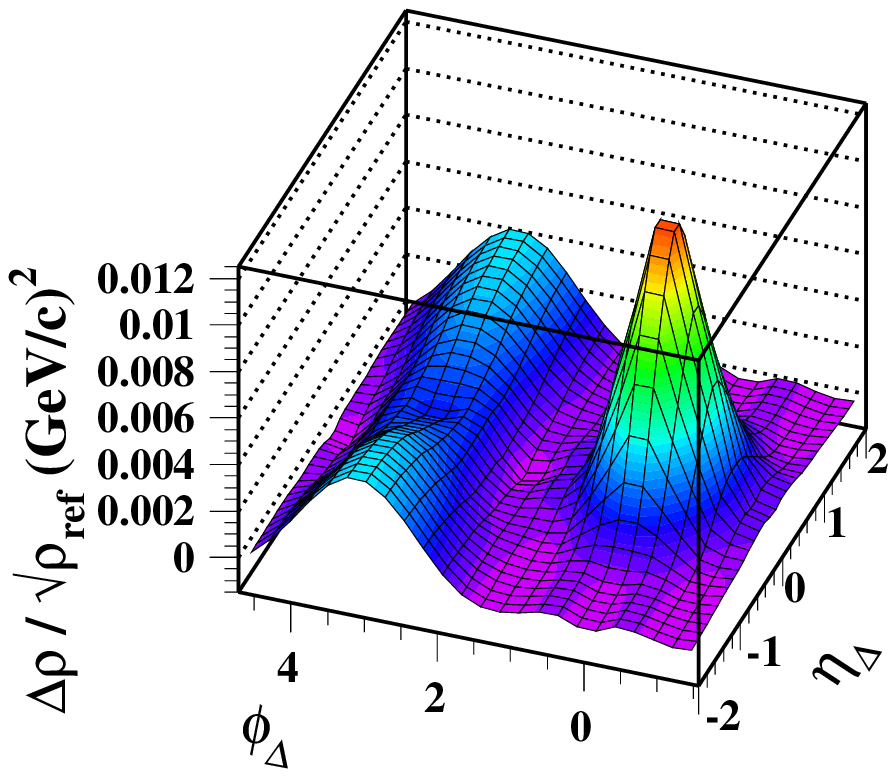} &
\includegraphics[keepaspectratio,width=1.65in]{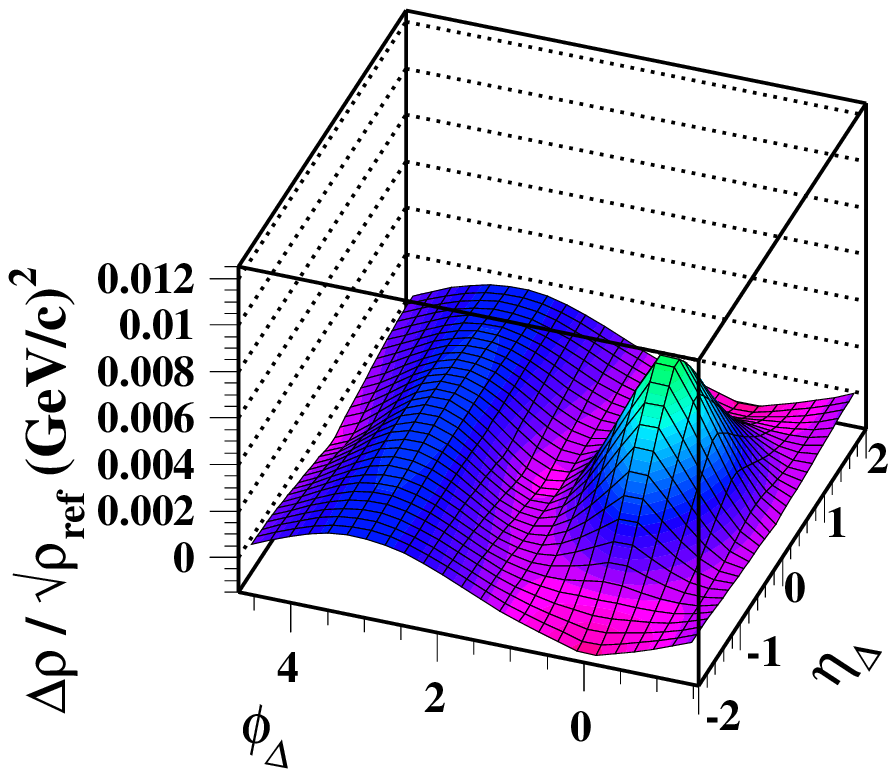}
\end{tabular}
\caption{Autocorrelations $\Delta \rho / \sqrt{\rho_{ref}}$ (GeV/c)$^2$ for Hijing quench-off (left panels) and quench-on (right panels) and for centralities 80-95\%  (top panels), 30-50\% (middle panels) and 0-15\% (bottom panels). \label{fig2}}
\end{figure}

Fig.~\ref{fig2} shows autocorrelations for three centralities (80-95\%, 30-50\% and 0-15\% from top to bottom) and two Hijing configurations: quench-off (left panels) and quench-on (right panels). All autocorrelations have similar features: a same-side ($|\phi_\Delta| < \pi / 2$) peak and an away-side ($|\phi_\Delta| > \pi / 2$) ridge. The same-side peak can be identified with fragments from {\em minimum-bias} partons (minijets), since no high-$p_t$ trigger condition was imposed. A related structure in RHIC data with high-$p_t$ trigger for angular or {\em number} (not $p_t$) correlations can be found in \cite{highptcorr}. The away-side ridge can be identified with momentum conservation for some combination of longitudinal string fragmentation and fragments from semi-hard parton scattering. There is some indication of a saddle structure or minimum at $\eta_\Delta = 0$ in the away-side ridge.

The variation of the near-side peak structure with centrality is limited. For quench-off collisions there is a modest increase in peak amplitude and small symmetric decrease in peak widths with increasing centrality. For quench-on collisions those trends are reversed. The amplitude falls and the peak widths increase, also varying symmetrically with increasing centrality.

\section{Errors}

The statistical errors for $\langle p_t \rangle$ fluctuation measurements as in Fig.~\ref{fig1} (left panels) are typically less than 0.001 (GeV/c)$^2$ for all scales and centralities. Autocorrelation errors have two components: statistical errors and image distortion due to smoothing. Statistical errors on the image are found by inverting statistical errors on the fluctuation data. The structure of that error distribution is typically simple and can be characterized by a single number for each autocorrelation. The smoothing error is estimated by passing inferred image ${\bf I}_\alpha$ a second time through the integration/inversion process to obtain ${\bf I}_{\alpha\alpha}$~\cite{inverse}. The difference between the two image distributions provides an upper limit on the smoothing distortion (because some of that change could include further reduction of statistical noise).
A typical image-error distribution is shown in Fig.~\ref{fig3} (right panel). The maxima typically occur near regions of large gradient in the autocorrelation. The per-bin systematic error on autocorrelation distributions is typically a few percent of maximum autocorrelation values.

\section{Model Fits}

\begin{figure}[t]
\includegraphics[keepaspectratio,width=3.3in]{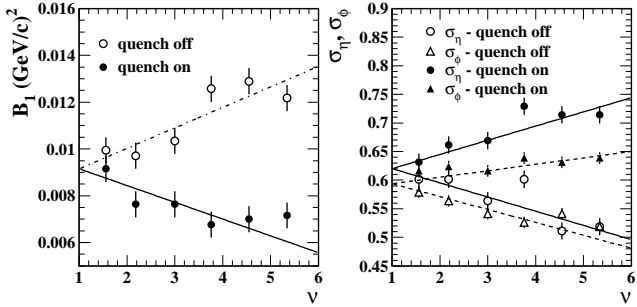}
\caption{Left panel: Correlation amplitudes {\em vs} centrality measure $\nu$ (mean participant path length) for near-side peaks and for quench-on collisions (dots) or quench-off collisions (triangles). Right panel: widths for near-side peaks on $\eta_\Delta$ (dots) and $\phi_\Delta$ (triangles) and for quench-on collisions (solid symbols) and quench-off collisions (open symbols). \label{fig4}}
\end{figure}

We fit the near-side structure for six centralities and two MC states with the model function in Eq.~(\ref{fit}): an offset plus a single 2D peak 
\bea \label{fit}
F &=& B_0 + B_1\, e^{-\left\{\left|\frac{\eta_\Delta}{\sqrt{2}\, \sigma_{\eta}}\right|^{\tau_\eta}+\left|\frac{\phi_\Delta}{\sqrt{2}\, \sigma_{\phi}}\right|^{\tau_\phi}\right\}}.
\eea
The residuals from fitted peak structures are typically less than 5\% of the near-side peak amplitude. Resulting peak amplitudes and widths are plotted in Fig.~\ref{fig4} {\em vs} centrality measure $\nu$, which estimates mean participant path length in number of encountered nucleons~\cite{nu}. The lines are drawn to guide the eye, with the constraint that quench-off and quench-on parameters should extrapolate to the same value for $\nu = 1$ (p-p collisions). Variation of the peak shape (exponents $\tau$) for these $p_t$ correlations was explored. The exponent values $\tau_\eta, \tau_\phi = 1.7 \pm 0.1$ provided the best description for all quench-on centralities. The peak shapes for quench-off data were closer to gaussian, but with a significant nonstatistical spread in the best-fit values: $\tau_\eta, \tau_\phi = 2.0 \pm 0.2$. There was no systematic difference between exponents for $\eta_\Delta$ and $\phi_\Delta$ directions.



\section{Discussion}

Interpretation of nonstatistical $\langle p_t \rangle$ fluctuations in terms of velocity/temperature correlations on $(\eta,\phi)$ is described in~\cite{mtxmt}. The present analysis provides direct access to the correlations suggested in~\cite{mtxmt}, in the form of $p_t$ autocorrelations obtained from inversion of  $\langle p_t \rangle$ fluctuation scale dependence. Those autocorrelations measure {\em covariances} of $\langle p_t \rangle $ fluctuations in pairs of small $\eta,\phi$ bins separated by $(\eta_\Delta,\phi_\Delta)$. That statement describes the second line of Eq.~(\ref{eq1}). Nonzero $p_t$ correlations indicate that particle $p_t$ is drawn event-wise from an {\em effective parent} distribution which deviates, as a function of $(\eta,\phi)$ and  {\em differently in each event}, from the inclusive $p_t$ distribution. $p_t$ autocorrelations reveal the two-point correlation structure on $(\eta,\phi)$ of {\em global properties} of the effective parent $p_t$ distribution, such as event-wise local transverse velocity and temperature. The near-side peaks can be interpreted as minijets, seen for the first time as {\em local velocity structures} in the pre-hadronic medium.

The modest centrality dependence of the same-side peak for both quench-off and default Hijing simulations is notable. The centrality dependence of Hijing $\langle p_t \rangle$ fluctuations measured in~\cite{QT} disagreed with RHIC measurements reported in~\cite{ptprl}. In that earlier Hijing study $\langle p_t \rangle$ fluctuations showed little centrality dependence, whereas $\langle p_t \rangle$ fluctuations in RHIC data were observed to be strongly centrality dependent~\cite{ptprl}. In the present Hijing study $p_t$ autocorrelations provide differential confirmation of trends in the previous study, and also confirm our previous interpretation of Hijing $\langle p_t \rangle$ fluctuations as mainly due to minijets. We might expect minijet production, at least in the more peripheral A-A collisions, to follow binary-collision scaling: an increase nearly proportional to participant-nucleon mean path length (estimated by centrality parameter $\nu$). The near-side peak amplitude in the $p_t$ autocorrelation should then be appoximately proportional to $\nu$, at least for quench-off collisions, but the observed increase is actually much slower. The autocorrelation changes in response to jet quenching are closer to expectations. When quenching is invoked the near-side peak amplitude is indeed significantly reduced relative to quench-off, and to an extent proportional to $\nu$. Peak widths on both difference variables increase, symmetrically and by only $10\,$-$\,20\%$. The linear dependence on $\nu$ of jet-quenching effects in this model is expected. Although $\nu$ estimates mean path length for initial-state participant nucleons it is also proportional to the average distance through the medium traversed by an outgoing parton.

\section{Summary}

In this analysis we have measured the scale dependence of $\langle p_t \rangle$ fluctuations for simulated Au-Au collisions in several centrality classes obtained from the Hijing-1.37 Monte Carlo. We have inverted those scale distributions to obtain $p_t$ autocorrelation distributions. We observe distinct minijet structures as near-side symmetric peaks which can be interpreted as velocity/temperature correlations on $(\eta,\phi)$. The $p_t$ autocorrelation structures, especially the near-side peak and its response to the jet-quenching mechanism in Hijing, agree qualitatively with perturbative-QCD expectations for semi-hard parton scattering followed by subsequent gluon bremsstrahlung depending on parton path length in the medium. The detailed centrality dependence of minijet correlations in Hijing seems inconsistent with expectations for binary-collision scaling. The near-side peak amplitude for quench-off collisions, representing hard parton scattering, increases by only 30\% from p-p ($\nu = 1$) to central Au-Au, where one would expect a 4$\times$ increase for binary-collision scaling. The near-side peak width variations with centrality are small and symmetric on $(\eta_\Delta,\phi_\Delta)$. These Hijing simulation results provide a valuable reference for heavy ion collision data from RHIC. We have introduced a novel analysis technique which reveals in heavy ion collisions new transverse-momentum correlation structures related to parton scattering, in-medium parton energy loss and parton fragmentation.

This work was supported in part by the Office of Science of the U.S. DoE under grant DE-FG03-97ER41020.


\end{document}